\documentclass[a4paper, amsfonts, amssymb, amsmath, reprint, showkeys, nofootinbib, twoside, superscriptaddress]{revtex4-2}
\usepackage[english]{babel}
\usepackage[utf8]{inputenc}

\usepackage[pdftex, pdftitle={Article}, pdfauthor={Author}]{hyperref}

\usepackage{graphicx}
\usepackage{dcolumn}
\usepackage{bm}
\usepackage{siunitx}
\usepackage{pstricks}
\usepackage{comment}
\usepackage{booktabs}
\usepackage{setspace}
\usepackage{ragged2e}
\usepackage{upgreek}

\usepackage{braket}
\usepackage{siunitx}
\usepackage{xcolor}  
\usepackage{gensymb}
\DeclareSIUnit\bar{bar}
\DeclareSIUnit\cps{cps}
\usepackage{textgreek}

\begin{document}

\DeclareSIUnit\bar{bar}
\DeclareSIUnit\cps{cps}
\title{A tunable, continuous-wave 130-mW laser at 213 nm}

\author{Lukas Möller}
\affiliation{Physikalisches Institut, Rheinische Friedrich-Wilhelms-Universität Bonn, Germany}
\author{Stephan Hannig}
\affiliation{Agile Optic GmbH, Braunschweig, Germany}
\author{Stefan Truppe}
\affiliation{Center for Cold Matter, Blackett Laboratory, Imperial College London, United Kingdom}
\author{Simon Stellmer}
\email{stellmer@uni-bonn.de}
\affiliation{Physikalisches Institut, Rheinische Friedrich-Wilhelms-Universität Bonn, Germany}

\begin{abstract}
We present a tunable, continuous-wave (CW) laser system emitting at 213 nm, based on the frequency quadrupling of a single Ti:sapph laser. The setup features two sequential, cavity-enhanced second-harmonic generation (SHG) stages. The first stage uses an LBO crystal and achieves a conversion efficiency of over 80\%, yielding up to 3.3 W at 426 nm. The second stage employs a Brewster-cut BBO crystal with an elliptical beam waist to mitigate UV-induced degradation, producing up to 130 mW of deep UV light. The system demonstrates stable operation over several hours and is well suited for applications in atomic physics, spectroscopy, and materials science.
\end{abstract}

\maketitle

High-power laser sources in the deep ultra-violet (DUV) wavelength range are an essential tool for many industrial and scientific applications, including semiconductor wafer inspection, interferometric lithography, angle-resolved photoemission spectroscopy (ARPES) and laser-cooling of certain atoms and molecules \cite{Brueck, Boschini, Kaneda, Shaw}. Many of these applications benefit from or require the use of single-frequency, continuous-wave (CW) lasers. Since there are no readily available direct CW sources in the DUV, such laser systems often consist of high-power laser sources in the near-infrared (NIR) wavelength range and nonlinear materials capable of converting the NIR light into the UV range. A common approach uses two consecutive second-harmonic generation (SHG) stages to achieve frequency quadrupling, employing enhancement cavities to improve conversion efficiency. Although sum-frequency generation (SFG) schemes can achieve higher powers at shorter wavelengths \cite{Sakuma}, they require multiple high-power lasers and introduce additional complexity and reduced tunability, making them less suitable for frequency-sensitive applications in atomic and molecular physics.

The conversion into the UV range requires a non-linear crystal that is transparent at both the fundamental and converted wavelengths and supports phase matching between the two wavelengths. Two well-established crystals for this process are CsLiB$_6$O$_{10}$ (CLBO) and $\beta$-BaB$_2$O$_4$ (BBO). While CLBO offers high conversion efficiency and low walk-off, it cannot phase-match below $\SI{234}{\nano \meter}$ \cite{Kaneda}, meaning that conversion to lower wavelengths needs to rely on BBO or the afore mentioned SFG processes. In recent years watt-level powers in the range of $\lesssim \SI{250}{\nano \meter}$ \cite{Burkley} and over $\SI{500}{\milli \watt}$ in the $\lesssim \SI{230}{\nano \meter}$ range \cite{Kaneda} have been demonstrated using these frequency quadrupled NIR lasers. However, for shorter wavelengths ($<\SI{220}{\nano \meter}$) it becomes more challenging to achieve high output powers, due to lower available fundamental powers from corresponding NIR sources and reduced efficiencies of the SHG process in BBO.

In this Letter, we present a $\SI{213}{\nano \meter}$ DUV laser system based on the fourth harmonic generation (FHG) of a titanium-sapphire laser, using two consecutive cavity-enhanced SHG stages. Our system uses an enhancement cavity with an elliptical waist to decrease light intensity on the crystal, while maintaining efficient conversion. The maximum output power achieved is $\SI{130}{\milli \watt}$. The wavelength of $\SI{213}{\nano \meter}$  corresponds to the fifth harmonic of Nd:YAG lasers, where pulsed systems are popular in material inspection and ARPES studies. In addition, the laser cooling transition of zinc, which has been proposed as a possible frequency standard in an optical clock \cite{Wang}, is located at $\SI{213.6}{\nano \meter}$ and thus provides an additional motivation to develop tunable narrow-linewidth lasers in this wavelength range.

\begin{figure}[htpb]
	\centering
	\includegraphics[width=\linewidth]{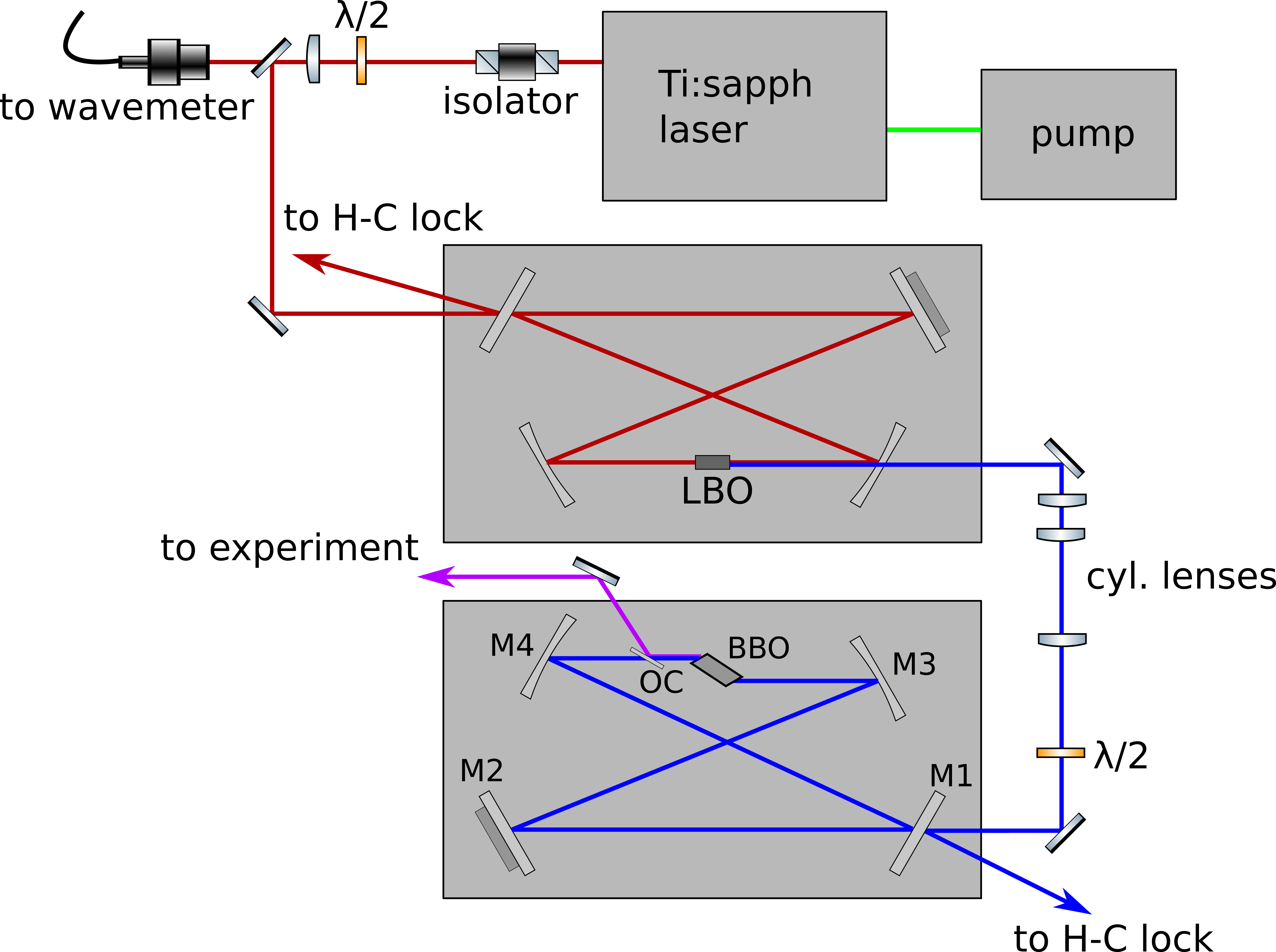}
	\caption{Schematic representation of the laser system. OC: Output coupler. H-C: Hänsch-Couillaud. The cavity mirrors of the second SHG stage are referred to as M1-4. M3 and M4 are cylindrical mirrors.}
	\label{fig:system}
\end{figure}

Titanium:sapphire (Ti:sapph) lasers offer high power, narrow linewidths, and exceptionally broad tunability, making them a natural choice for nonlinear frequency conversion into the deep UV. In our system, a commercial Ti:sapph laser (MSquared SolsTis), pumped by an $\SI{18}{\watt}$ frequency-doubled Nd:YAG laser, delivers up to $\SI{4}{\watt}$ at $\SI{854.4}{\nano \meter}$ and provides excellent beam quality and spectral purity.
The free-running frequency drift of about $\SI{10}{\mega \hertz}$ per hour is sufficient for most atomic and molecular physics applications. However, the system requires regular realignment to compensate for gradual degradation in power and mode quality over 1 to 2 days, likely caused by mechanical drift in the laser cavity. It is also highly sensitive to optical feedback, necessitating a high-extinction (60\,dB) optical isolator.

The first SHG stage (Agile Optic) uses an LBO crystal to frequency-double the $\SI{854.4}{\nano \meter}$ light to $\SI{427.2}{\nano \meter}$. The crystal is placed in a bow-tie enhancement cavity consisting of two planar and two concave mirrors, forming a $\sim \SI{25}{\micro \meter}$ waist at the crystal. One of the planar mirrors is mounted on a piezo actuator for scanning or active stabilization. The cavity is locked to the laser using the Hänsch–Couillaud technique \cite{HC}, with the reflected light from the input coupler providing an error signal for a PID controller that drives the piezo.

Mode matching into the cavity is achieved with a single lens, reaching a coupling efficiency exceeding $90\%$. The finesse of the cavity is about 300. The produced $\SI{427.2}{\nano \meter}$ light is coupled out of the cavity through one of the curved mirrors. The cavity produces up to $\SI{3.3}{\watt}$ of $\SI{427.2}{\nano \meter}$ light, which corresponds to an external conversion efficiency of above $80\%$. 

The $\SI{213.6}{\nano \meter}$ light is generated in a 10 mm long Brewster-cut BBO crystal. Similarly to the first SHG stage, the crystal is placed in a bow-tie cavity (Agile Optic). Instead of circular concave mirrors, it uses cylindrical ones to generate an elliptical waist at the position of the BBO crystal. When used for the generation of DUV light, BBO crystals are prone to UV damage from the light generated within the crystal. The elliptical waist helps mitigate this issue by distributing the light within the crystal over a larger area, while maintaining a tight focus in the direction of walk-off \cite{Preißler}. Elliptical focusing has previously enabled stable watt-level UV generation at $\SI{257}{\nano \meter}$ without degradation of the BBO crystal \cite{Gumm}. The use of cylindrical mirrors increases the number of degrees of freedom needed to optimize cavity performance, since their rotation will influence the beam waist at the position of the crystal and the shape of the cavity mode.

Just like the first SHG stage, one of the mirrors is mounted to a piezo to enable frequency scanning or active frequency stabilization of the cavity. Due to the cylindrical mirrors, the blue light that is to be coupled into the cavity needs to be shaped into an elliptical beam to ensure proper mode-matching to the cavity. This is achieved by a system of three cylindrical lenses, resulting in $\sim 70\%$ incoupling efficiency. The BBO crystal is mounted to a kinematic stage, which allows for the fine adjustment of the crystal position and phase-matching angle.

Since BBO is hygroscopic, the cavity is sealed and continuously purged with filtered, dry air. The crystal is heated to $\SI{145}{\celsius}$ to enhance its resistance to UV-induced degradation \cite{Takachiho}. Heating also improves reproducibility, as we observed consistent performance across BBO crystals from different suppliers when operated above $\SI{100}{\celsius}$. This is in contrast to previous reports of substantial variation at room temperature \cite{Kaneda}.

\begin{figure}[t]
	\centering
    \includegraphics[width=0.8\linewidth]{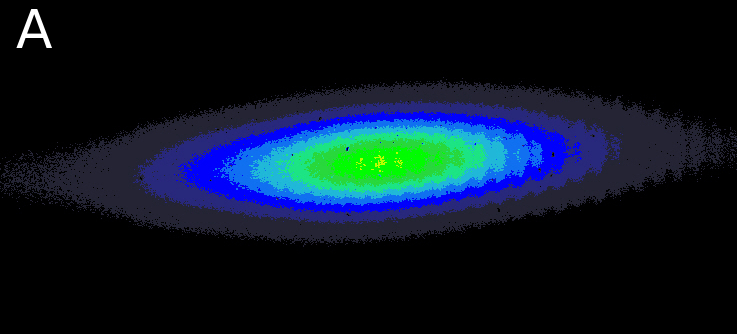}
    \includegraphics[width=0.8\linewidth]{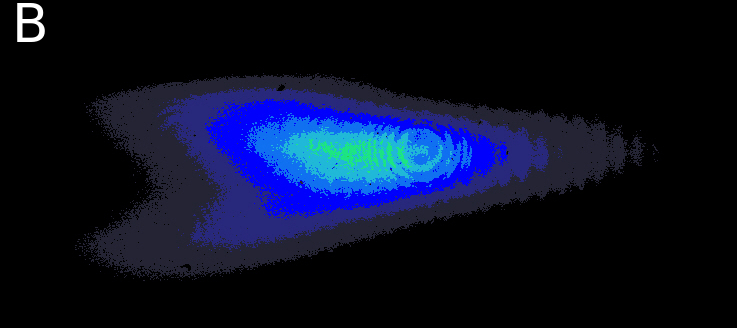}
	\caption{Distortion of the cavity mode, caused by thermal lensing at the output coupler. A: Undisturbed mode during scanning of the cavity; B: Disturbed mode during locking.}
	\label{fig:modes}
\end{figure}

In order to prevent the UV light from damaging the cavity mirrors, it is coupled out of the cavity by placing a dichroic mirror (OC in Fig. \ref{fig:system}), with high reflectivity (R>$99.7\%$) for $\SI{213.6}{\nano \meter}$ and high transmission (T>$99.8\%$) for $\SI{427.2}{\nano \meter}$, between the crystal and the following cavity mirror (M4 in Fig. \ref{fig:system}). We found that this output coupler significantly impacts the performance of the cavity when it is locked. We assume that the high circulating power inside the cavity causes the output coupler to heat up locally, resulting in thermal lensing, which distorts the mode circulating in the cavity. The mode circulating in the cavity can be observed by recording the beam profile of the light leaking through one of the cavity mirrors. Fig. \ref{fig:modes} shows a comparison between the cavity modes while the cavity is scanned (no thermal lensing) A and while it is locked B. The severity of this effect depends on the material and thickness of the output coupler.

Additionally, the high circulating power also heats up the BBO and, by doing so, changes the phase-matching angle required to generate UV light. As a consequence of these thermal effects, some of the cavity parameters, like the alignment of the cavity mirrors or the position of the BBO, need to be optimized while the cavity is locked.
Since the cavity is sealed, the UV output passes through an additional fused silica window imprinted with a random anti-reflection (RAR) nanostructure to maximize transmission.

\begin{figure}[t]
	\centering
    \includegraphics[width=\linewidth]{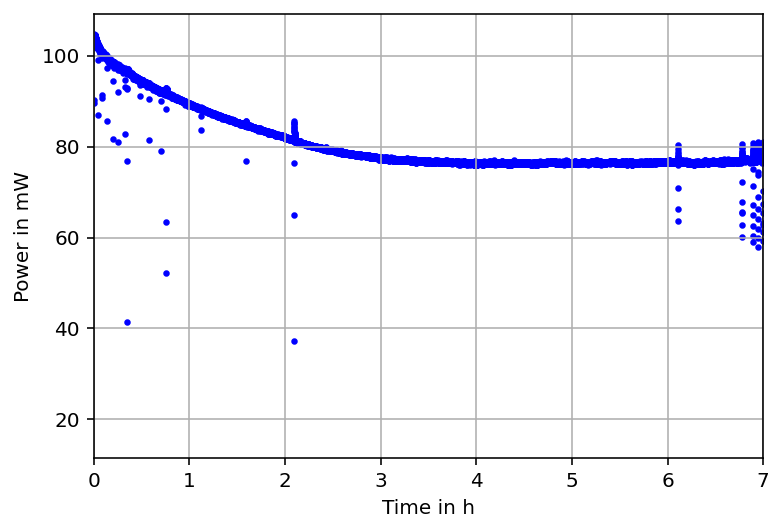}
	\caption{Measurement of the laser's output power over 7 hours. The occasional sudden drops in power are caused by relocking of the cavity. The apparent increase in power after relocking is caused by the detector.}
	\label{fig:power}
\end{figure}

With this system, we are regularly able to generate powers above $\SI{100}{\milli \watt}$. The maximum power generated by this system is $\SI{130}{\milli \watt}$. Fig. \ref{fig:power} shows a measurement of the output power of the laser over 7 hours of continuous operation. The power drops to $\sim \SI{80}{\milli \watt}$ within a few hours and settles there for the remainder of the 7 hours. This drop in power is likely due to heating of the BBO as well as the formation of temperature gradients in the BBO, causing a non-uniform phase-matching condition across its length.

For operation over multiple days, the power drops continuously. We attribute this to instability of incoupling of the $\SI{427.2}{\nano \meter}$ light into the second SHG stage and possibly damage to the output coupler. The output power of the first SHG stage remains constant for multiple days. After multiple days of operation, the full output power can be recovered by slightly realigning the mirrors and crystal position of the second SHG stage.
We have not observed any clear signs of irreversible damage to the crystal after cumulative operation for more than $\SI{150}{\hour}$.
We have taken advantage of the tunability of the Ti:sapph to change the DUV wavelength by more than $\SI{1}{\nano \meter}$ without a significant decrease in output power or conversion efficiency.

Seven hours of stable operation make the system suitable for applications in atomic physics. Here, slight changes in the wavelength would allow to address important wavelengths such as twice the Lyman-$\beta$ transition at 205\,nm. Sum frequency generation with a powerful, narrow-linewidth and frequency-stable NIR laser would allow to shift the wavelength into the range of 180\,nm. In condensed matter physics, CW DUV sources are advantageous for ARPES due to the suppression of space charge effects compared to pulsed excitation \cite{Zhou}. The present system was specifically developed for laser cooling of zinc, where previous experiments were limited by short DUV exposure times and insufficient optical power \cite{roeser}.

In the future, we plan to further increase stability and output power of the system by replacing the Ti:sapph and first SHG stage with an intracavity doubled vertical-external-cavity surface-emitting-laser (VECSEL). Such systems are already well-established and widely used for frequency conversion \cite{Guina}.

\begin{figure}[t]
	\centering
    \includegraphics[width=\linewidth]{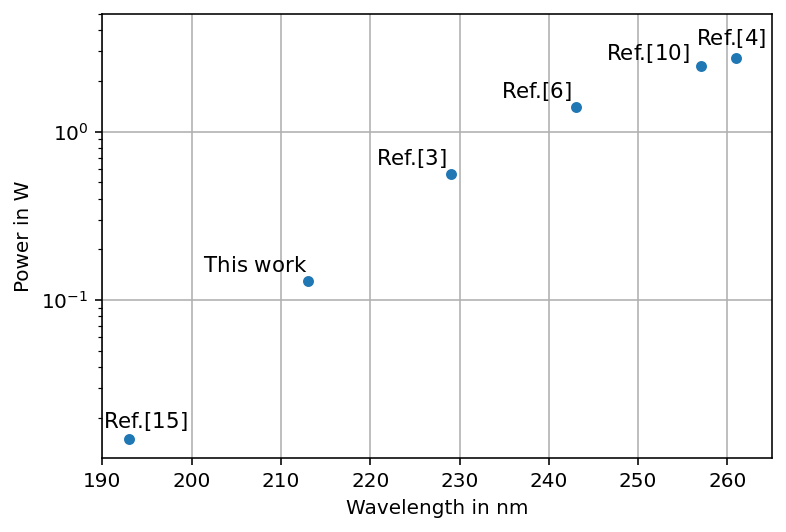}
	\caption{Comparison of output powers achieved at different UV wavelengths, using frequency-quadrupled IR/NIR lasers \cite{Kaneda, Shaw, Burkley, Gumm, Scholz}.}
	\label{fig:comparison}
\end{figure}

While $\SI{130}{\milli \watt}$ may appear modest compared to powers achieved at longer wavelengths, a decrease in output power is expected as the target wavelength shortens. This trend is illustrated in Fig. \ref{fig:comparison}, which summarizes reported output powers from systems based on tunable, single-frequency IR or NIR sources with two sequential SHG stages. The data show a clear decline in achievable power at shorter wavelengths, and our result at $\SI{130}{\milli \watt}$ fits well within this trend.

\begin{figure}[htpb]
	\centering
    \includegraphics[width=\linewidth]{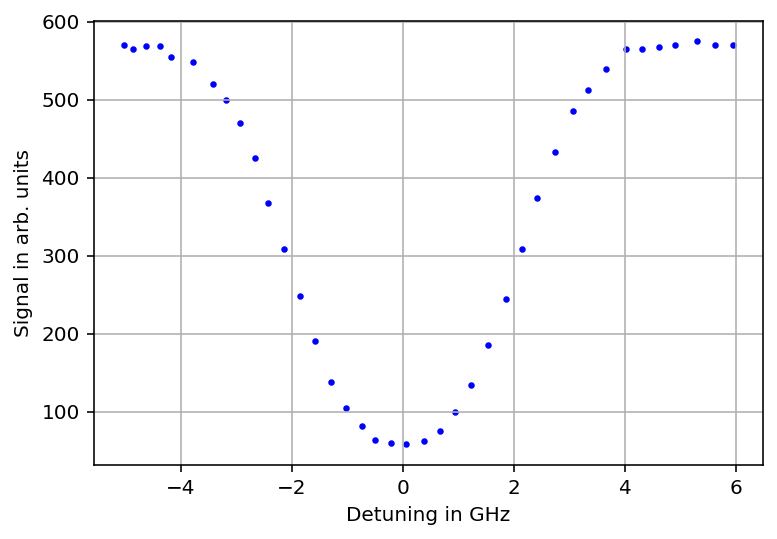}
	\caption{Doppler-broadened spectroscopy signal for the $^1S_0 \xrightarrow{} {^1P_1}$ in zinc, around a center frequency of $\nu_0=\SI{1401392}{\giga \hertz}$.}
	\label{fig:doppler}
\end{figure}

To illustrate the feasibility of using this system for applications in atomic physics, we perform spectroscopy on the $^1S_0 \xrightarrow{} {^1P_1}$ of zinc at $\SI{213.6}{\nano \meter}$. The resulting Doppler-broadened signal is shown in Fig. \ref{fig:doppler} and demonstrates the good tunability and  stability of the system.

In summary, we present a tuneable CW laser source at $\SI{213.6}{\nano \meter}$, suitable for multiple applications like the laser-cooling of zinc or ARPES. Our system is based on a single Ti:sapph laser that is frequency quadrupled in two consecutive SHG stages. The system is capable of generating stable output for more than 7 hours. A maximum power of $\gtrsim \SI{130}{\milli \watt}$ has been achieved. We assume that the difference between the maximum and the equilibrated output power is caused by thermal effects, such as the formation of temperature gradients within the BBO crystal. The stability of the system appears to be limited by the mechanical stability of the optics and the output coupler of the second cavity. Key design features, such as elliptical focusing, dry air purging, and high-temperature operation, successfully mitigate degradation of the BBO crystal, as confirmed by stable performance over more than 150 hours of operation.

{Funding - }
We acknowledge funding from Deutsche Forschungsgemeinschaft DFG through grants INST 217/978-1 FUGG and 496941189, as well as through the Cluster of Excellence "ML4Q" (EXC 2004/1 – 390534769), from the European Research Council (ERC) under the European Union’s Horizon 2020 Research and Innovation Programme (Grant Agreement No. 757386 "quMercury"), and from the European Commission through project 101080164 "UVQuanT".

{Acknowledgments - }
We thank D.~Röser and M.~Vöhringer for early experimental work and J.~Domarkas of Eksma Optics for providing dedicated optics. We thank the entire team of the UVQuanT consortium for inspiring discussions and technical advice, especially Sid C.~Wright of FHI Berlin for close collaboration and critical reading of the manuscript.

{Disclosures - }
The authors declare no conflicts of interest.

{Data Availability Statement - }
Data underlying the results presented in this paper are not publicly available at this time but may be obtained from the authors upon reasonable request.


\bibliography{bib}

\end{document}